\documentclass[10pt,twocolumn,journal]{IEEEtran}

\usepackage{lettrine}
\usepackage{amsmath}
\usepackage{cases}
\usepackage{bm}
\usepackage{txfonts}
\usepackage{amssymb}
\usepackage{cite}
\usepackage{graphicx}
\usepackage{psfrag}
\usepackage{url}
\usepackage{stfloats}
\usepackage{algorithm}
\usepackage{algorithmic}        
\usepackage{multirow}
\usepackage{color}
\usepackage{romannum}
\usepackage{lipsum}
\usepackage{stfloats}
\usepackage[utf8]{inputenc}
\usepackage[skip=2pt,font=scriptsize]{caption}
\usepackage[skip=2pt,font=scriptsize]{subcaption}
\usepackage{tabularx}
\usepackage{booktabs}

\ifCLASSINFOpdf

\else
\fi

\begin{document}
\pagenumbering{arabic}
\title{Generative Model for Joint Resource Management in Multi-Cell Multi-Carrier NOMA Networks}

\author{ Elhadj Moustapha Diallo
\thanks{\IEEEauthorrefmark{1}Corresponding author: Elhadj Moustapha Diallo  (L2@cqupt.edu.cn)}
\thanks{E. M. Diallo is with School of Communications and Information Engineering, Chongqing University of Posts and Telecommunications, Chongqing,
P. R. China, 400065.}
}

\maketitle

\begin{abstract}
In this work, we design a generative artificial intelligence(GAI) -based framework for joint resource allocation, beamforming, and power allocation in multi-cell multi-carrier non-orthogonal multiple access (NOMA) networks. We formulate the proposed problem as sum rate maximization problem. Next, we design a novel multi-task transformer (MTT) framework to handle the problem in real-time. To provide the necessary training set, we consider simplified but powerful mathematical techniques from the literature. Then, we train and test the proposed MTT. We perform simulation to evaluate the efficiency of the proposed MTT and compare its performance with the mathematical baseline.
\end{abstract}


\section{Introduction}
The integration of non-orthogonal multiple access (NOMA) into multi-cell networks represents a paradigm shift in wireless communication, aimed at fulfilling the escalating demand for enhanced spectral efficiency and accommodating the exponential growth in number of devices \cite{bb1}. Central to the optimization of NOMA systems is the joint application of user scheduling \cite{bb2,bb3}, beamforming \cite{bb4}, and power allocation \cite{bb5}-three interdependent techniques that collectively unlock NOMA's potential to revolutionize network performance for the 6G. The joint optimization of these problems not only addresses the complexities inherent in managing dense, heterogeneous networks but also sets a new standard for achieving unprecedented levels of network capacity, reliability, and energy efficiency \cite{bb6}.

Deep learning-based resource allocation in NOMA systems have received great attention from researchers \cite{bb7,bb8}. Unlike the traditional optimization methods, deep learning frameworks are very suitable for application where real-time/semi-real-time is crucial. In this realm, the advent of generative artificial intelligence (GAI) has heralded a new era in the optimization of wireless systems \cite{bb9,bb10,bb11}. However, particularly in addressing the intricate challenges posed by NOMA in multi-cell environments is not yet well addressed. Specifically, the joint optimization of user scheduling, beamforming, and power allocation in such settings since it present a complex problem space, characterized by dynamic network conditions and the need for real-time decision-making. GAI, with its ability to model and simulate complex systems, offers a groundbreaking approach to solving these challenges, ushering in significant advancements in network performance, efficiency, and reliability \cite{bb12}. In this work, we aim at leveraging the advantages of GAI to jointly handle user scheduling, beamforming, and power allocation in multi-cell multi-carrier NOMA networks. Our contributions in this work can be summarized as follows:
\begin{itemize}
  \item We consider a downlink multi-cell multi-carrier NOMA network. Aiming at maximizing the sum rate, we formulate the optimization problem with the consideration of users scheduling, power budget, and quality-of-service (QoS) constraints. The formulate problem is NP-hard which is difficult to handle with reasonable complexity using traditional optimization techniques. Therefore, we resort into deep learning.
  \item We design a novel multi-task transformer (MTT) framework to handle the problem. The proposed MTT architecture can capture local radio information and shadowing effects in radio propagation to accommodate irregularly distributed practical observations in proposed network architecture. Moreover, the proposed MTT performs joint user scheduling, beamforming, and power allocation in real-time.
  \item We provide experimental comparisons with state-of-the-art methods in terms of both accuracy and complexity. Our experimental results demonstrate the strength of our proposed framework in capturing both global model information and local shadowing features, thereby facilitating radio-map-assisted resource management for 5G and B5G wireless systems.
\end{itemize}
The remainder of this paper is organized as follows: In Section II, we discuss the system model and problem formulation. In Section III, we provide mathematical solution for the formulated problem. In Section IV, we introduce the proposed MTT framework. Section V, investigates the simulation results and discussion. In Section VI, we give the conclusions and the future works.
\section{System Model and Problem Formulation}
We consider a multi-cell multi-carrier interference network includes $M$ base stations (BSs) serving $K$ users through $S$ subcarriers. Let ${\mathcal{M}} = \left\{ {1,2,....,M} \right\}$,  ${\mathcal{K}} = \left\{ {1,2,....,K} \right\}$, and $\mathcal{S} = \left\{1,2,...,S\right\}$ denote the set of BSs, users, and subcarriers, respectively. We assume multiple-input multiple-output (MISO) where each BS is equipped with $N$ antenna while each user is equipped with single antenna. We define the binary variable $\rho_{m,k,s}\in \left\{0,1\right\}$ to denote the user scheduling in which $\rho_{m,k,s} = 1$ if the user $k$ is associated with the BS $m$ via the subcarrier $s$, and $\rho_{m,k,s} = 0$, otherwise. The received signal at the user $k$ associated with the BS $m$ via the subcarrier $s$ can be given as
\begin{equation}\label{eqn1}
\begin{aligned}
{y_{m,k,s}} &= \underbrace {{\bf{h}}_{m,k,s}^H{\rho _{m,k,s}}{{\bf{w}}_{m,k,s}}\sqrt {{p_{m,k,s}}} {s_{m,k,s}}}_{{\text{deired signal}}}\\
 & + \underbrace {\sum\limits_{j \ne k}^K {{\bf{h}}_{m,k,s}^H{\rho _{m,k,s}}{{\bf{w}}_{m,j,s}}\sqrt {{p_{m,j,s}}} {s_{m,j,s}}} }_{{\text{Intra - cell interference}}}\\
 &+ \underbrace {\sum\limits_{n \ne m} {{\bf{h}}_{n,k,s}^H\sum\limits_{k' = 1}^K {{\rho _{n,k',s}}{{\bf{w}}_{n,k',s}}\sqrt {{p_{n,k',s}}} {s_{n,k',s}}} } }_{{\text{Inter - cell interference}}} + \underbrace {{n_{m,k,s}}}_{{\text{Noise}}},
\end{aligned}
\end{equation}

where ${{\bf{h}}_{m,k,s}} \in {\mathbb{C}^{N \times 1}}$ is the channel coefficient, $E\left[ {{{\left| {{s_{m,k,s}}} \right|}^2}} \right] = 1$ , ${{\bf{w}}_{m,k,s}} \in {\mathbb{C}^{N \times 1}}$ with $\left\| {{{\bf{w}}_{m,k,s}}} \right\| = 1$ is the beamforming vector, and   is the additive white Gaussian noise (AWGN). After performing successive interference cancelation (SIC), the signal-to-interference-plus-noise ratio (SINR) of the user $k$ associated with the BS $m$ through the subcarrier $s$ can be written as in \eqref{eqn2} on top of page \pageref{eqn2}.
\begin{figure*}
\centering
\begin{equation}\label{eqn2}
{\gamma _{m,k,s}} = \frac{{{{\left| {{\rho _{m,k,s}}{\bf{h}}_{m,k,s}^H{{\bf{w}}_{m,k,s}}} \right|}^2}{p_{m,k,s}}}}{{\sum\limits_{j = 1,j \ne k}^K {{{\left| {{\rho _{m,k,s}}{\bf{h}}_{m,k,s}^H{{\bf{w}}_{m,j,s}}} \right|}^2}{p_{m,j,s}}}  + \sum\limits_{n \ne m} {\sum\limits_{k' = 1}^K {{{\left| {{\rho _{n,k',s}}{\bf{h}}_{n,k,s}^H{{\bf{w}}_{n,k',s}}} \right|}^2}{p_{n,k',s}}} }  + \sigma _{m,k,s}^2}},
\end{equation}
\end{figure*}

The achievable rate of the user $k$ associated with the BS $m$ via the subcarrier $s$ is given as
\begin{equation}\label{eqn3}
{r_{m,k,s}} = B{\log _2}\left( {1 + {\gamma _{m,k,s}}} \right),
\end{equation}
We aim at maximizing the network sum rate by optimizing the user scheduling, beamforming and power allocation. The optimization problem can be formulated as

\begin{subequations}\label{eqn4:main}
\begin{align}
& \mathop {\max }\limits_{\rho ,{\bf{w}},p} & & \sum\limits_{m = 1}^M {\sum\limits_{k = 1}^K {\sum\limits_{s = 1}^S {{r_{m,k,s}}} } } & & \tag{\ref{eqn4:main}}\\
&\text{subject to}&& \sum\limits_{m = 1}^M {\sum\limits_{s = 1}^S {{\rho _{m,k,s}}{p_{m,k,s}}} }  \ge 0,\forall k,\label{eqn4:a}\\
&                 && \sum\limits_{k = 1}^K {\sum\limits_{s = 1}^S {{\rho _{m,k,s}}{p_{m,k,s}}} }  \le {P_m},\forall m, \label{eqn4:b}\\
&                 && {r_{m,k,s}} \ge {r_{\min }},\forall k,\label{eqn4:c}\\
&                 && {\rho _{m,k,s}} \in \left\{ {0,1} \right\},\forall m,\forall s,\label{eqn4:d}\\
&                 && \sum\limits_{m = 1}^M {\sum\limits_{s = 1}^S {{\rho _{m,k,s}}} }  \le 1,\forall m,\forall s,\label{eqn4:e}
\end{align}
\end{subequations}
The problem in \eqref{eqn4:main} mixed-integer non-linear programming (MINLP) which is nonconvex and NP hard problem \cite{bb13}. In the following section, we provide a generative framework to handle the problem in \eqref{eqn4:main}.

\section{Joint Optimization of User Scheduling, Beamforming and Power Allocation in NOMA Networks}

\subsection{User Scheduling Subproblem}
The user scheduling subproblem can be written as follows
\begin{subequations}\label{eqn5:main}
\begin{align}
& \mathop {\max }\limits_{\rho} & & \sum\limits_{m = 1}^M {\sum\limits_{k = 1}^K {\sum\limits_{s = 1}^S {{r_{m,k,s}}} } } & & \tag{\ref{eqn5:main}}\\
&\text{subject to}&&\eqref{eqn4:d}, \eqref{eqn4:e},\notag
\end{align}
\end{subequations}
Using the method in \cite{bb17}, we handle the user scheduling with fixed beamforming and power allocation. Towards that, first we introduce the auxiliary variable $\phi_k = \sum\limits_{m = 1}^M {\sum\limits_{s = 1}^S {{\rho _{m,k,s}}} }$. Thus, the user scheduling subproblem in \eqref{eqn5:main} is rewritten as below
\begin{equation}\label{eqn6}
\mathop {\max }\limits_{0\le \phi_k \le 1} \sum\limits_{m = 1}^M {\sum\limits_{k = 1}^K {\sum\limits_{s = 1}^S {{r_{m,k,s}}} } },
\end{equation}
The solution of the problem in \eqref{eqn6} is given as follows
\begin{equation}\label{eqn7}
  \begin{aligned} {\phi _{k}}=&\begin{cases} 0,&{\mathrm {if}}\mathop {\arg\max }\limits_{m \in {\mathcal{M}}} \sum\limits_{m = 1}^M {\sum\limits_{k = 1}^K {\sum\limits_{s = 1}^S {{r_{m,k,s}}} } } \le 0 \\ \\ \mathop {\arg\max }\limits_{m \in {\mathcal{M}}} \sum\limits_{m = 1}^M {\sum\limits_{k = 1}^K {\sum\limits_{s = 1}^S {{r_{m,k,s}}} } }, &{}{\mathrm {otherwise}} \end{cases}
\end{aligned}
\end{equation}
\begin{equation}\label{eqn8}
  \begin{aligned} {\phi _{k}}=&\begin{cases} 0,&{\mathrm {if}}\mathop {\arg\max }\limits_{s \in {\mathcal{S}}} \sum\limits_{m = 1}^M {\sum\limits_{k = 1}^K {\sum\limits_{s = 1}^S {{r_{m,k,s}}} } } \le 0 \\ \\ \mathop {\arg\max }\limits_{s \in {\mathcal{S}}} \sum\limits_{m = 1}^M {\sum\limits_{k = 1}^K {\sum\limits_{s = 1}^S {{r_{m,k,s}}} } }, &{}{\mathrm {otherwise}} \end{cases}
\end{aligned}
\end{equation}
The solutions in \eqref{eqn7} and \eqref{eqn8} are the user-BS association and subcarrier allocation, respectively. These solutions provide the necessary training dataset for the user scheduling part.
\subsection{Beamforming Subproblem}
Given the user scheduling and the power allocation, the beamforming subproblem can be written as
\begin{subequations}\label{eqn9:main}
\begin{align}
& \mathop {\max }\limits_{{\bf{w}}} & & \sum\limits_{m = 1}^M {\sum\limits_{k = 1}^K {\sum\limits_{s = 1}^S {{r_{m,k,s}}} } } & & \tag{\ref{eqn9:main}}\\
&\text{subject to}&&\eqref{eqn4:c},\notag
\end{align}
\end{subequations}
To handle the problem in \eqref{eqn9:main}, we follow the procedure in \cite{bb4} to obtain the beamforming. First, we apply semi-definite relaxation (SDR) to reformulate the problem as follows
\begin{subequations}\label{eqn10:main}
\begin{align}
& \mathop {\max }\limits_{{\bf{w}}} & & \sum\limits_{m = 1}^M {\sum\limits_{k = 1}^K {\sum\limits_{s = 1}^S {{r_{m,k,s}\left( {\bf{w}}\right)}} } } & & \tag{\ref{eqn10:main}}\\
&\text{subject to}&&{r_{m,k,s}}\left( {\bf{w}}\right) \ge {r_{\min }},\label{eqn10:a}\\
&                 && {\bf{w}}_{m,k,s} \succeq {\mathbf{0}},\label{eqn10:b}
\end{align}
\end{subequations}
where ${r_{m,k,s}}\left( {\bf{w}}\right)$ is given as in \eqref{eqn11} on top of page \pageref{eqn11}.
\begin{figure*}
  \begin{equation}\label{eqn11}
{r_{m,k,s}}\left( {\bf{w}} \right) = B{\log _2}\left( {1 + \frac{{\left( {{{\bf{h}}_{m,k,s}}{{\bf{W}}_{m,k,s}}{\bf{h}}_{m,k,s}^H} \right){p_{m,k,s}}}}{{\sum\limits_{j = 1,j \ne k}^K {\left( {{{\bf{h}}_{m,k,s}}{{\bf{W}}_{m,j,s}}{\bf{h}}_{m,k,s}^H} \right){p_{m,j,s}}}  + \sum\limits_{n \ne m} {\sum\limits_{k' = 1}^K {\left( {{{\bf{h}}_{n,k,s}}{{\bf{W}}_{n,k',s}}{\bf{h}}_{n,k,s}^H} \right){p_{n,k',s}}} }  + \sigma _{m,k,s}^2}}} \right),
\end{equation}
\end{figure*}
Problem \eqref{eqn10:main} can be relaxed using the S-Lemma as in \cite{bb14,bb15}. Under reasonable conditions, this relaxation is tight and the problem in \eqref{eqn10:main} can be solved using SDR.

\begin{figure*}
\centering
  \includegraphics[width= 14 cm,height=7cm]{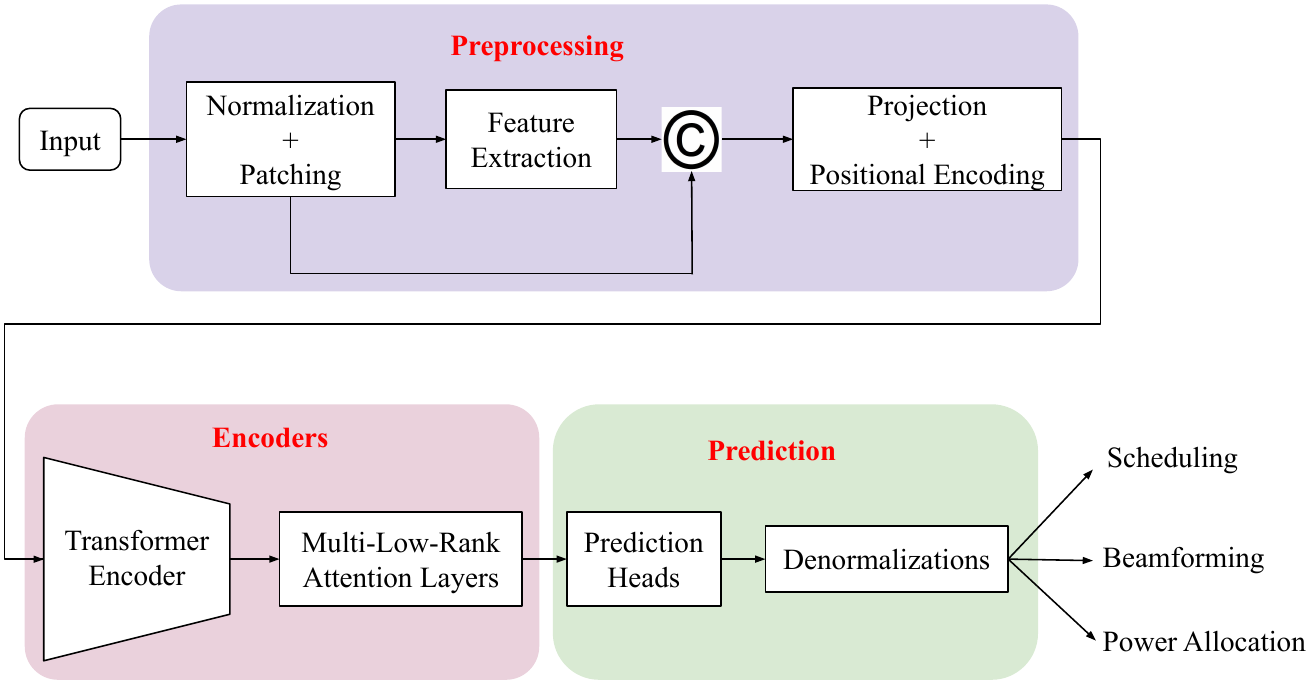}
  \caption{Structure of the proposed MTT framework.}
  \label{ovll}
\end{figure*}
\subsection{Power Allocation Subproblem}
Fixing the user scheduling and the beamforming, we write the power allocation subproblem as follows
\begin{subequations}\label{eqn12:main}
\begin{align}
& \mathop {\max }\limits_{p} & & \sum\limits_{m = 1}^M {\sum\limits_{k = 1}^K {\sum\limits_{s = 1}^S {{r_{m,k,s}}} } } & & \tag{\ref{eqn12:main}}\\
&\text{subject to}&&\eqref{eqn4:a}, \eqref{eqn4:b}, \eqref{eqn4:c},\notag
\end{align}
\end{subequations}
we apply successive pseudo-convex approximation (SPCA) \cite{bb18,bb19} on the objective function in \eqref{eqn12:main} and the constraint \eqref{eqn4:c}. Thus, the approximated ${\tilde r_{m,k,s}}\left( {{p_{m,k,s}};{\bf{p}}^{\left( t \right)}} \right)$ is given as
\begin{equation}\label{eqn13}
\begin{aligned}
{{\tilde r}_{m,k,s}}\left( {{p_{m,k,s}};{{\bf{p}}^{\left( t \right)}}} \right) &= {r_{m,k,s}}\left( {{p_{m,k,s}};{\bf{p}}_{m, - k,s}^{\left( t \right)}} \right)\\
 &+ \sum\limits_{k' \ne k}^K {\left( {{p_{m,k,s}};p_{m, - k,s}^{\left( t \right)}} \right){\nabla _{{p_{m,k,s}}}}{r_{m,k',s}}\left( {{{\bf{p}}^{\left( t \right)}}} \right)} ,
\end{aligned}
\end{equation}
The power allocation ${p_{m,k,s}}$ can be obtained as fixed-point expression after applying Lagrange relaxation and applying Karush-Kuhn-Tucker (KKT) conditions.

\section{Multi-Task Transformer for Joint Resource Management in Multi-Cell Multi-Carrier NOMA Networks}
In this section, provide the architecture of the proposed multi-task transformer (MTT) to perform real-time joint user scheduling, beamforming, and power allocation in multi-cell multi-carrier NOMA networks. The over architecture is illustrated in Fig. \ref{ovll}.

\subsection{MTT Description}
Given the input which consists of the channel coefficients, number of BSs, number of subcarriers, $r_{min}$, and $P_m$, the proposed MTT performs normalization and patching to the input. Next, an inception-residual block is employed to perform feature extraction. Next, we have combination of projection and position embeddings allows MTT to effectively process multivariate INPUT data by ensuring that the model not only understands the content of the data but also the order and timing of observations, which are critical for prediction tasks. Projection helps standardize the representations of different features into a consistent format that can be efficiently processed by the subsequent neural network layers. The projection involves a dense (fully connected) layer to linearly transform the input features into a higher-dimensional space. The transformation is parameterized by learnable weights, which are optimized during training to capture essential features relevant for forecasting. On the other hand, position embeddings are crucial in models like Transformers that are inherently permutation-invariant (i.e., they do not consider the order of input sequences naturally). These embeddings provide the model with information about the relative or absolute position of the samples in the input sequence.

The core of the MTT model includes a transformer encoder that processes the INPUT embedding derived from previous representations. The output of the transformer encoder is fed into specialized multi-low-rank attention layer to efficiently capture cross-dimensional dependencies, enhancing the model's ability to handle channel-wise interactions without significant increases in computational demands. After processing through the transformer encoder and multi-low-rank attention layers, the transformed data is fed into a prediction head, which is responsible for producing the final forecasting output. The Prediction heads apply generalized linear units (GELU) followed by dropout layer and linear layers for the case of beamforming and power allocation. For the case of user scheduling, the GELU and linear layers are replaced by sigmoid layer.

\subsection{System Setup and Training Procedure of MTT}
We use the mathematical solution in Section III to generate the necessary training dataset. We generate 115000 sample which is divided into 90\% for training and 10\% for validation. The simulation parameters for the case of wireless network are provided in Section V. The batch size is set to 128 and the training epochs is 300 while Adam optimizer is employed for the optimization. To prevent overfitting, the dropout is set to 0.05.

The neural network framework is implemented in PyTorch 2.0 and Python 3.8 platform. The computer specifications are: Intel(R) Core(TM) i9-10900K CPU @ 3.70GHz, NVIDIA Quadro RTX 5000 graphic card, and 64 GB memory.
\section{Simulation Results}
In this section, we introduce the simulation results and performance evaluation for the proposed models. The cell radius is 500 m. The users are randomly and uniformly distributed in the cells. The subchannel frequency is 250 kHz and the carrier frequency is 2.4 GHz. The noise power spectral density is -80 dBm/Hz and the noise figure is 7dB. The simulation parameters are deemed default as shown in TABLE \ref{sim} unless stated otherwise.
\begin{table}
\centering
\caption{Simulation Parameters}
\label{sim}
\begin{tabular}{|l|l|}
  \hline
  \textbf{Parameter} & \textbf{Value} \\
  \hline
   Number of users per BS &	12\\
   \hline
   Number of BSs &	4\\
   \hline
   Number of antennas $N$ & 2\\
   \hline
   Fast-Fading &	$Z \sim \mathcal{CN}\left(0,1\right)$\\
   \hline
  Path loss exponent &	3.7\\
  \hline
   Number of BS antennas &	8\\
   \hline
  $B$ &	250 kHz\\
  \hline
  Carrier frequency	& 2.4 GHz\\
  \hline
  Circuit power &	20 dBm\\
  \hline
\end{tabular}
\end{table}

\begin{figure}[!ht]
\centering
\includegraphics[width=3.5in]{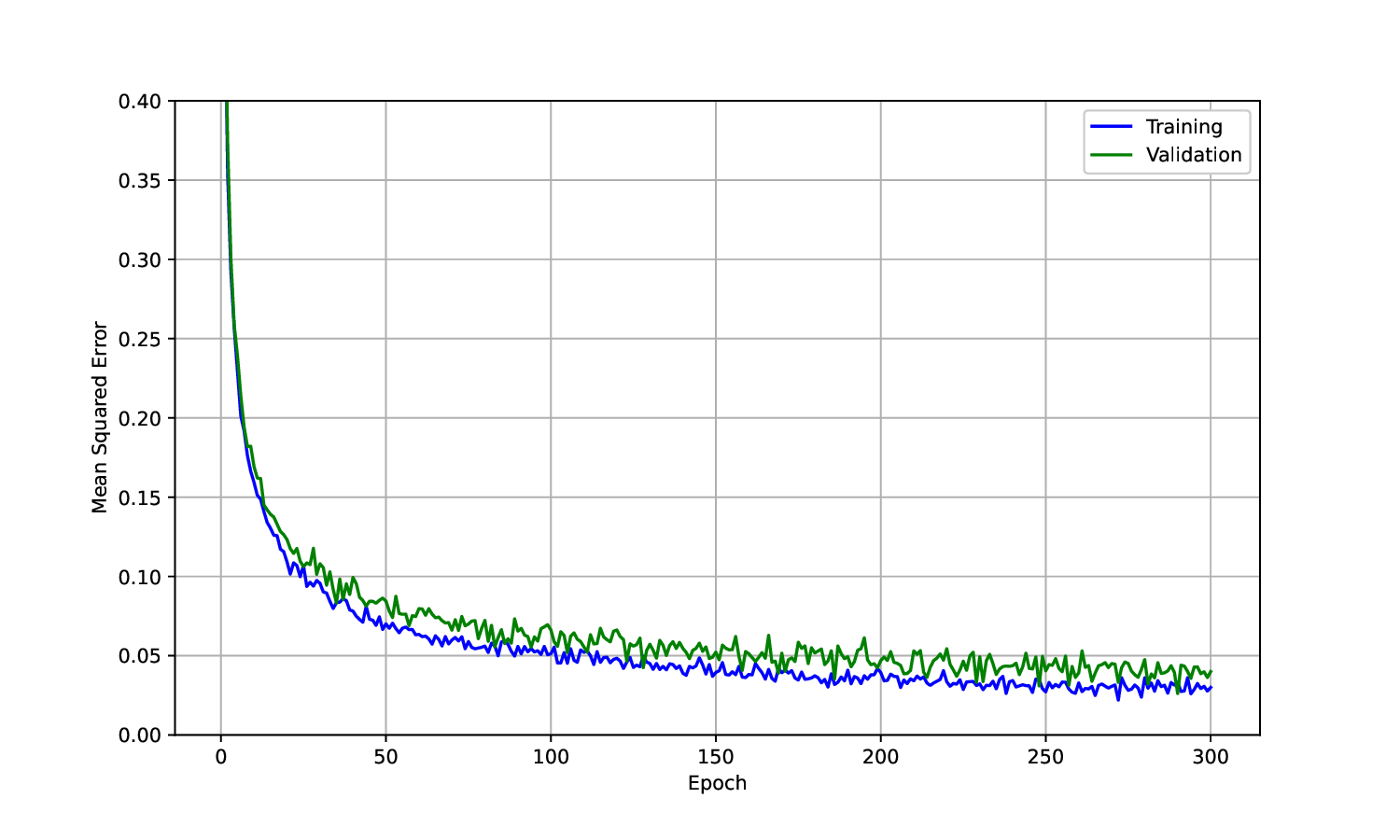}
\caption{Training and validation accuracy of the proposed model.}
\label{acc}
\end{figure}

The performance in terms of training and validation accuracy is presented in Fig. \ref{acc}. The number of the subcarrier is set as $S = {K \mathord{\left/
 {\vphantom {K 2}} \right.
 \kern-\nulldelimiterspace} 2}$. As we can see, the mean square error decreases with the increase in the number of epochs until it converges after 275 epochs.

MTT demonstrated robust performance, suggesting that its architecture is well-suited to handle the complexities and variabilities inherent in wireless network optimization.

The design of MTT with a multi-low-rank attention mechanism, deep feature extraction, with projection and positional embedding allows for efficient training. This architecture enables the model to capture significant features both globally and locally without incurring the high computational costs typically associated with deep transformer models. Moreover, the linear complexity with respect to the input sequence length, independent of its duration, suggests that the model converges faster than traditional transformer models, which often struggle with longer sequences due to their quadratic computational complexity.

\begin{figure}[!ht]
\centering
\includegraphics[width=3.5in]{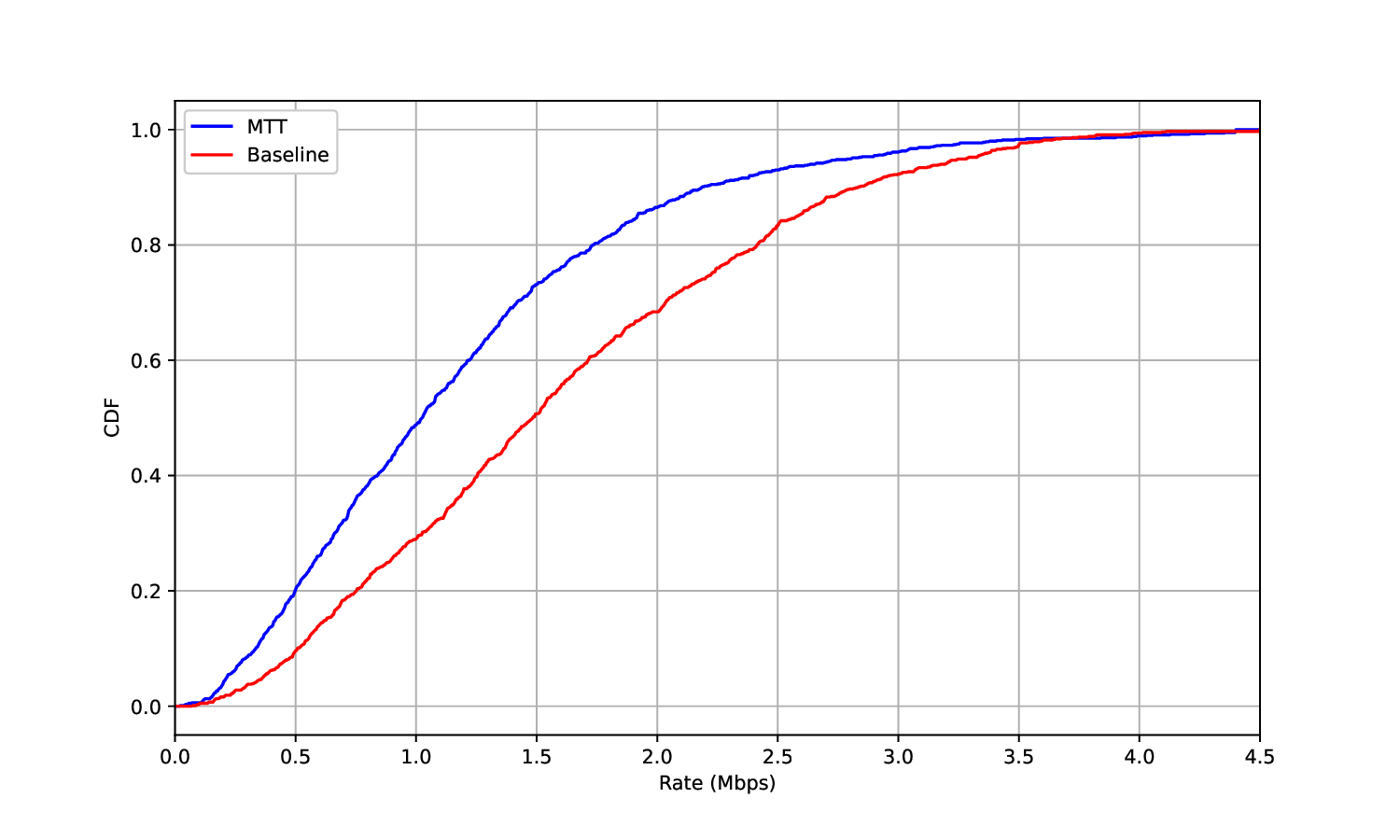}
\caption{Performance comparison between the proposed framework and the baseline.}
\label{cdf}
\end{figure}

Experiments demonstrated that MTT not only converges quickly but also requires significantly less memory and computational resources compared to other high-performance models. This efficiency is critical when deploying models in resource-constrained environments or where real-time forecasting is needed. Unlike the traditional optimization methods such as the baseline, the proposed MTT can perform in real-time and with comparable performance to the baseline as shown in Fig. \ref{cdf}.

\section{Conclusions and Future Works}
In this work, we considered the joint resource management in multi-cell multi-carrier NOMA networks. Aiming at maximizing the sum rate, we formulate an optimization problem with power budget and QoS constraints. To handle the problem, we focus on optimizing the user scheduling, beamforming, and power allocation. Towards this, first we provide mathematical solution to provide the necessary data and baseline for comparison. Second, we design a novel transformer architecture (namely MTT) to handle the problem in real-time. The simulation results show the efficiency of the proposed MTT in optimizing the problem. As a future direction, we shall consider the application of the proposed MTT in dynamic nature such as low earth orbit (LEO) satellite-based networks, and unmanned aerial vehicle (UAV)-aided networks.

\ifCLASSOPTIONcaptionsoff
  \newpage
\fi

\end{document}